\documentclass{IEEEtran}

\usepackage{amsmath,amssymb,amsthm,braket,dsfont,subfig,graphicx,hyperref}
\usepackage[noadjust]{cite}

\input{qcircuit}

\DeclareMathOperator{\Tr}{Tr}
\DeclareMathOperator{\openone}{\mathds{1}}

\newtheorem{lmm}{Lemma}
\newtheorem{thm}{Theorem}
\newtheorem{cor}{Corollary}

\begin{document}

\title{Guesswork of a quantum ensemble}

\author{Michele  Dall'Arno\thanks{M. Dall'Arno  is with  the
    Yukawa   Institute   for  Theoretical   Physics,   Kyoto
    University, Sakyo-ku,  Kyoto, 606-8502, Japan,  and with
    the  Faculty  of  Education   and  Integrated  Arts  and
    Sciences,   Waseda   University,   Shinjuku-ku,   Tokyo,
    169-8050,                 Japan                (e--mail:
    dallarno.michele@yukawa.kyoto-u.ac.jp)},       Francesco
  Buscemi\thanks{F.   Buscemi  is  with  the  Department  of
    Mathematical    Informatics,    Graduate    School    of
    Informatics,  Nagoya   University,  Chikusa-ku,  Nagoya,
    464-8601, Japan (e--mail: buscemi@nagoya-u.jp)}, Takeshi
  Koshiba,~\IEEEmembership{Member,~IEEE}\thanks{T.  Koshiba
    is with the Faculty of Education and Integrated Arts and
    Sciences,   Waseda   University,   Shinjuku-ku,   Tokyo,
    169-8050,                 Japan                (e--mail:
    tkoshiba@waseda.jp)}\thanks{YITP-20-161}}

\maketitle

\begin{abstract}
  The guesswork of a quantum ensemble quantifies the minimum
  number of guesses needed in average to correctly guess the
  state of the ensemble, when  only one state can be queried
  at a  time.  Here, we  derive analytical solutions  of the
  guesswork problem  subject to a finite  set of conditions,
  including the  analytical solution for any  qubit ensemble
  with  uniform   probability  distribution.    As  explicit
  examples, we  compute the guesswork for  any qubit regular
  polygonal and polyhedral ensemble.
\end{abstract}

\section{Introduction}
\label{sec:introduction}

We consider a communication  scenario involving two parties,
Alice and  Bob.  An ensemble $\boldsymbol{\rho}$  of quantum
states with labels in a set $\mathcal{M}$ is given and known
to both parties.  At each round,  Alice picks a label $m \in
\mathcal{M}$  with  probability  $\Tr[\boldsymbol{\rho}(m)]$
and     hands      state     $\Tr[\boldsymbol{\rho}(m)]^{-1}
\boldsymbol{\rho}(m)$  over to  Bob. Bob  aims at  correctly
guessing label  $m$ being  allowed to  query one  element of
$\mathcal{M}$  at a  time, until  his query  is correct,  at
which point the round is over. The cost function incurred by
Bob is the average number of guesses, or \textit{guesswork},
until he correctly guesses  $m$. Bob's most general strategy
consists    of    performing     a    quantum    measurement
$\boldsymbol{\pi}$  outputing an  element $\mathbf{n}$  from
the   set  $\mathcal{N}_{\mathcal{M}}$   of  numberings   of
$\mathcal{M}$ and querying the  elements of $\mathcal{M}$ in
the order  specified by $\mathbf{n}$.  Hence,  the guesswork
is  given  by  the  occurence  of  label  $m$  in  numbering
$\mathbf{n}$,  averaged  over  all  numberings.   Using  the
formalism~\cite{Wil17} of quantum circuits,  the setup is as
follows:
\begin{align}
  \label{eq:setup}
  \Qcircuit  @C=1em  @R=1em  {\lstick{m \in  \mathcal{M}}  &
    \prepareC{\boldsymbol{\rho} \left( m  \right)} \cw & \qw
    &     \ustick{\mathcal{H}}      \qw     &      \qw     &
    \measureD{\boldsymbol{\pi} \left( \mathbf{n} \right )} &
    \rstick{\mathbf{n} \in \mathcal{N}_{\mathcal{M}}.} \cw}
\end{align}

The  guesswork has  been extensively  studied for  classical
ensembles~\cite{Mas94,  Ari96,  AM98a, AM98b,  Pli98,  MS04,
  Sun07, HS10,  CD13, SV18,  Sas18}, but only  very recently
tackled   for    quantum   ensembles~\cite{CCWF15,   HKDW20,
  DBK21}. While previous works  focused on the derivation of
entropic  bounds,  our  aim  is instead  the  derivation  of
analytical      solutions.        Our      main      result,
Theorem~\ref{thm:helstrom}, provides  an analytical solution
subject  to  a finite  set  of  conditions.  In  particular,
Corollary~\ref{cor:qubit}  provides the  analytical solution
for   any   qubit    ensemble   with   uniform   probability
distribution,  thus disproving  the conjecture~\cite{CCWF15}
that  analytical   solutions  exist  only  for   binary  and
symmetric    ensembles.     As   explicit    examples,    in
Corollaries~\ref{cor:polygons}   and~\ref{cor:polyhedra}  we
explicitly  compute  the  minimum  guesswork  of  any  qubit
regular  polygonal  and polyhedral  ensebles,  respectively.
This proves  a conjecture~\cite{HKDW20} on the  guesswork of
the square qubit ensemble.

\section{Formalization}
\label{sec:formalization}

In  this section  we define  the guesswork  problem. We  use
standard      results      from     quantum      information
theory~\cite{Wil17}.

First,   we   introduce   the    sets   of   ensembles   and
numbering-valued measurements  that appear  in the  setup of
Eq.~\eqref{eq:setup}.   For any  finite dimensional  Hilbert
space  $\mathcal{H}$,   we  denote  with   $\mathcal{L}_+  (
\mathcal{H} )$ the cone  of positive semi-definite operators
on  $\mathcal{H}$.  For  any  finite  set $\mathcal{M}$,  we
denote   with   $\mathcal{N}_{\mathcal{M}}$   the   set   of
numberings given by
\begin{align*}
  \mathcal{N}_{\mathcal{M}} := \left\{  \mathbf{n} : \left\{
  1,   \dots,  \left|   \mathcal{M}  \right|   \right\}  \to
  \mathcal{M} \Big| \mathbf{n} \textrm{ bijective} \right\}.
\end{align*}
We denote with  $\mathcal{E}( \mathcal{M}, \mathcal{H})$ the
set of ensembles given by
\begin{align*}
  \mathcal{E}  \left( \mathcal{M},  \mathcal{H} \right)  : =
  \left\{ \boldsymbol{\rho} :  \mathcal{M} \to \mathcal{L}_+
  \left( \mathcal{H} \right)  \Big| \sum_{m \in \mathcal{M}}
  \Tr \left[ \boldsymbol{\rho} \left(  m \right) \right] = 1
  \right\}.
\end{align*}
and   with  $\mathcal{P}(   \mathcal{N}_{\mathcal{M}},
\mathcal{H}  )$  the  set of  numbering-valued  measurements
given by
\begin{align*}
  \mathcal{P} \left(  \mathcal{N}_{\mathcal{M}}, \mathcal{H}
  \right)      :=       \left\{      \boldsymbol{\pi}      :
  \mathcal{N}_{\mathcal{M}}    \to   \mathcal{L}_+    \left(
  \mathcal{H}    \right)    \Big|    \sum_{\mathbf{n}    \in
    \mathcal{N}_{\mathcal{M}}}    \boldsymbol{\pi}    \left(
  \mathbf{n} \right) = \openone \right\}.
\end{align*}

Next,  we  introduce   the  probability  distributions  that
describe  the   setup  in  Eq.~\eqref{eq:setup}.    For  any
ensemble   $\boldsymbol{\rho}$   and  any   numbering-valued
measurement     $\boldsymbol{\pi}$,    we     denote    with
$p_{\boldsymbol{\rho},    \boldsymbol{\pi}}$    the    joint
probability    distribution    that     the    outcome    of
$\boldsymbol{\pi}$  is numbering  $\mathbf{n}$ and  that the
$t$-th guess is correct, that is  $\mathbf{n} ( t ) = m$. In
formula:
\begin{align*}
  p_{\boldsymbol{\rho},   \boldsymbol{\pi}}   \;  :   \;   &
  \mathcal{N}_{\mathcal{M}}   \times  \{   1   ,  \dots,   |
  \mathcal{M}  |  \}  \longrightarrow   [0,  1]\\  &  \left(
  \mathbf{n},    t    \right)   \longmapsto    \Tr    \left[
    \boldsymbol{\rho}  \left(  \mathbf{n} \left(  t  \right)
    \right)   \boldsymbol{\pi}  \left(   \mathbf{n}  \right)
    \right],
\end{align*}
for  any  $\boldsymbol{\rho} \in  \mathcal{E}  (\mathcal{M},
\mathcal{H})$  and  any $\boldsymbol{\pi}  \in  \mathcal{P}(
\mathcal{N}_{\mathcal{M}},  \mathcal{H})$.   We denote  with
$q_{\boldsymbol{\rho},  \boldsymbol{\pi}}$  the  probability
distribution  that the  $t$-th  guess  is correct,  obtained
marginalizing    the    joint    probability    distribution
$p_{\boldsymbol{\rho}, \boldsymbol{\pi}}$. In formula:
\begin{align*}
  q_{\boldsymbol{\rho}, \boldsymbol{\pi}} \;  : \; & \left\{
  1, \dots, | \mathcal{M}  | \right\} \longrightarrow \left[
    0,  1\right]\\  &  t  \longmapsto  \sum_{\mathbf{n}  \in
    \mathcal{N}_{\mathcal{M}}}         p_{\boldsymbol{\rho},
    \boldsymbol{\pi}} \left( \mathbf{n}, t \right),
\end{align*}
for  any  $\boldsymbol{\rho} \in  \mathcal{E}  (\mathcal{M},
\mathcal{H})$  and  any $\boldsymbol{\pi}  \in  \mathcal{P}(
\mathcal{N}_{\mathcal{M}},  \mathcal{H})$.

Finally, we  are in a  position to introduce  the guesswork.
The \textit{guesswork}  $G$ is  a function mapping  any pair
$(\boldsymbol{\rho},  \boldsymbol{\pi})$   of  ensemble  and
numbering-valued measurement  into the expectation  value of
the  number $t$  of guesses,  averaged with  the probability
distribution  $q_{\boldsymbol{\rho},  \boldsymbol{\pi}}$  of
correctness of the $t$-th guess. In formula:
\begin{align*}
   G \; : \; & \mathcal{E} (\mathcal{M}, \mathcal{H}) \times
   \mathcal{P}  (   \mathcal{N}_{\mathcal{M}},  \mathcal{H})
   \longrightarrow [1, \infty)\\ & \left( \boldsymbol{\rho},
     \boldsymbol{\pi}   \right)    \longmapsto   \sum_{t   =
       1}^{\left| \mathcal{M} \right|} q_{\boldsymbol{\rho},
       \boldsymbol{\pi}} \left( t \right) t.
\end{align*}
The  \textit{minimum  guesswork}   $G_{\textrm{min}}$  is  a
function mapping  any ensemble $\boldsymbol{\rho}$  into the
minimum over numbering-valued  measurements of the guesswork
$G$.  In formula:
\begin{align*}
  G_{\textrm{min}}  \;   :  \;   &  \mathcal{E}(\mathcal{M},
  \mathcal{H})     \longrightarrow    [1,     \infty)\\    &
    \boldsymbol{\rho} \longmapsto \min_{\boldsymbol{\pi} \in
      \mathcal{P}      \left(     \mathcal{N}_{\mathcal{M}},
      \mathcal{H}   \right)}  G   \left(  \boldsymbol{\rho},
    \boldsymbol{\pi} \right).
\end{align*}

\section{Main results}
\label{sec:results}

In this  section we  derive the  analytical solution  of the
guesswork  problem subject  to a  finite set  of conditions,
including  any  qubit   ensemble  with  uniform  probability
distribution.

In order  to state  our main result,  we need  the following
definitions.   For  any  finite  dimensional  Hilbert  space
$\mathcal{H}$, we denote with  $\mathcal{L} ( \mathcal{H} )$
the space  of Hermitian operators on  $\mathcal{H}$. For any
finite set $\mathcal{M}$ and any ensemble $\boldsymbol{\rho}
\in \mathcal{E}(\mathcal{M},  \mathcal{H})$, we  denote with
$E_{\boldsymbol{\rho}}   :   \mathcal{N}_{\mathcal{M}}   \to
\mathcal{L} ( \mathcal{H} )$ the map given by
\begin{align*}
  E_{\boldsymbol{\rho}}  \left(  \mathbf{n}   \right)  &  :=
  \sum_{t  = 1}^{\left|  \mathcal{M} \right|}  \left( 2  t -
  \left| \mathcal{M}  \right| - 1  \right) \boldsymbol{\rho}
  \left( \mathbf{n} \left( t \right) \right),
\end{align*}
for any $\mathbf{n}  \in \mathcal{N}_{\mathcal{M}}$. For any
numbering  $\mathbf{n}  \in  \mathcal{N}_{\mathcal{M}}$,  we
denote  with   $  \overline{  \mathbf{n}  }$   the  reversed
numbering. In formula:
\begin{align*}
  \overline{\mathbf{n}}  \left(  t   \right)  :=  \mathbf{n}
  \left( \left| \mathcal{M} \right| + 1 - t \right),
\end{align*}
for any $t  \in \{ 1, \dots, | \mathcal{M}  | \}$. We denote
with $\Pi_- ( \cdot )$ and  $\Pi_0 ( \cdot )$ the projectors
on the negative and null parts of $( \cdot )$, respectively.
We  denote   with  $\{  \boldsymbol{\pi}_{\boldsymbol{\rho},
  \mathbf{n}^*} \in \mathcal{P}( \mathcal{N}_{\mathcal{M}} )
\}_{\mathbf{n}^* \in  \mathcal{N}_{\mathcal{M}}}$ the family
of numbering-valued measurements given by
\begin{align*}
  \boldsymbol{\pi}_{\boldsymbol{\rho},  \mathbf{n}^*} \left(
  \mathbf{n} \right) :=
  \begin{cases}
    \left(   \Pi_-   +    \frac12   \Pi_0   \right)   \left(
    E_{\boldsymbol{\rho}} \left( \mathbf{n} \right) \right),
    &   \textrm{if  }   \mathbf{n}   \in  \{   \mathbf{n}^*,
    \overline{\mathbf{n}}^* \},\\ 0, & \textrm{otherwise},
  \end{cases}
\end{align*}
for   any   $\mathbf{n}^*,  \mathbf{n}   \in   \mathcal{N}_{
  \mathcal{M}}$.   It follows  from Lemma~\ref{lmm:gw}  that
the corresponding guesswork is given by
\begin{align}
  \label{eq:mingw}
  G                 \left(                \boldsymbol{\rho},
  \boldsymbol{\pi}_{\boldsymbol{\rho}, \mathbf{n}^*} \right)
  = \frac{\left| \mathcal{M} \right| + 1}2 - \frac12 \left\|
  E_{\boldsymbol{\rho}} \left( \mathbf{n}^* \right) \right\|_1,
\end{align}
for any $\mathbf{n}^* \in \mathcal{N}_{\mathcal{M}}$.

Upon  denoting  with  $|  \cdot |$  the  absolute  value  of
operator  $(  \cdot  )$,   the  following  theorem  provides
analytical  solutions  of   the  minimum  guesswork  problem
subject to a finite set of conditions.

\begin{thm}
  \label{thm:helstrom}
  For any  finite set $\mathcal{M}$, any  finite dimensional
  Hilbert    space   $\mathcal{H}$,    and   any    ensemble
  $\boldsymbol{\rho}   \in    \mathcal{E}   (   \mathcal{M},
  \mathcal{H}  )$, if  there exists  numbering $\mathbf{n}^*
  \in \mathcal{N} ( \mathcal{M} )$ such that
  \begin{align}
    \label{eq:condition}
    \left| E_{\boldsymbol{\rho}} \left( \mathbf{n}^* \right)
    \right|  \ge   E_{\boldsymbol{\rho}}  \left(  \mathbf{n}
    \right),
  \end{align}
  for any  $\mathbf{n} \in  \mathcal{N}_{\mathcal{M}}$, then
  numbering-valued                               measurement
  $\boldsymbol{\pi}_{\boldsymbol{\rho},   \mathbf{n}^*}  \in
  \mathcal{P}  (  \mathcal{N}_{\mathcal{M}}, \mathcal{H}  )$
  minimizes the guesswork, that is
  \begin{align*}
    G_{\textrm{min}}  \left( \boldsymbol{\rho}  \right) =  G
    \left(                                \boldsymbol{\rho},
    \boldsymbol{\pi}_{\boldsymbol{\rho},       \mathbf{n}^*}
    \right).
  \end{align*}
\end{thm}

We remark  that, while the  minimum guesswork problem  is by
definition an  optimization over a  \textit{continuous} set,
the   conditions  given   by  Eq.~\eqref{eq:condition}   are
\textit{finite}  in  number  and  hence can  be  checked  by
exhaustive  search.   If   they  hold,  Eq.~\eqref{eq:mingw}
provides the  analytical solution  of the  minimum guesswork
problem.

\begin{proof}
  Due to Lemma~\ref{lmm:gw} one has $G_{\textrm{min}} (
  \boldsymbol{\rho}   )   =  (|   \mathcal{M}   |   +  1   +
  x_{\boldsymbol{\rho}}) / 2$, where
  \begin{align*}
    x_{\boldsymbol{\rho}}   :=  \min_{\boldsymbol{\pi}   \in
      \mathcal{P} \left(  \mathcal{N}_{\mathcal{M}} \right)}
    \sum_{\mathbf{n}   \in  \mathcal{N}_{\mathcal{M}}}   \Tr
    \left[  E_{\boldsymbol{\rho}} \left(  \mathbf{n} \right)
      \frac{  \boldsymbol{\pi} \left(  \mathbf{n} \right)  -
        \boldsymbol{\pi}     \left(    \overline{\mathbf{n}}
        \right) }2 \right].
  \end{align*}
  Since   for  any   $\boldsymbol{\pi}  \in   \mathcal{P}  (
  \mathcal{N}_{\mathcal{M}} )$  the sum is lower  bounded by
  its minimum term, one has
  \begin{align*}
    x_{\boldsymbol{\rho}}   \ge   y_{\boldsymbol{\rho}}   :=
    \min_{\substack{\boldsymbol{\pi} \in  \mathcal{P} \left(
        \mathcal{N}      \left(     \mathcal{M}      \right)
        \right)\\ \mathbf{n} \in \mathcal{N}_{\mathcal{M}}}}
    \Tr   \left[  E_{\boldsymbol{\rho}}   \left(  \mathbf{n}
      \right)  \frac{   \boldsymbol{\pi}  \left(  \mathbf{n}
        \right)       -        \boldsymbol{\pi}       \left(
        \overline{\mathbf{n}} \right) }2 \right].
  \end{align*}
  Using  Lemma~\ref{lmm:majorization1}, for  any $\mathbf{n}
  \in    \mathcal{N}_{\mathcal{M}}$    the   minimum    over
  $\boldsymbol{\pi}         \in        \mathcal{P}         (
  \mathcal{N}_{\mathcal{M}} )$ can be computed leading to
  \begin{align*}
    y_{\boldsymbol{\rho}}    =   -    \max_{\mathbf{n}   \in
      \mathcal{N}_{\mathcal{M}}}                     \left\|
    E_{\boldsymbol{\rho}}    \left(    \mathbf{n}    \right)
    \right\|_1.
  \end{align*}
  Using                            Eq.~\eqref{eq:condition},
  Lemma~\ref{lmm:majorization2},          and          again
  $E_{\boldsymbol{\rho}}      (      \mathbf{n}     )      =
  -E_{\boldsymbol{\rho}} (  \overline{ \mathbf{n} }  )$, the
  maximum  over  $\mathbf{n} \in  \mathcal{N}_{\mathcal{M}}$
  can be computed leading to
  \begin{align*}
    y_{\boldsymbol{\rho}} =  - \left\| E_{\boldsymbol{\rho}}
    \left( \mathbf{n}^* \right) \right\|_1.
  \end{align*}
  Since           $G          (           \boldsymbol{\rho},
  \boldsymbol{\pi}_{\boldsymbol{\rho},\mathbf{n}^*}  ) =  (|
  \mathcal{M}  |  + 1  +  y_{\boldsymbol{\rho}})  / 2$,  the
  statement follows.
\end{proof}

The following corollary provides  the analytical solution of
the minimum  guesswork problem  for any qubit  ensemble with
uniform probability distribution.

\begin{cor}
  \label{cor:qubit}
  For  any finite  set  $\mathcal{M}$,  any two  dimensional
  Hilbert    space   $\mathcal{H}$,    and   any    ensemble
  $\boldsymbol{\rho}   \in    \mathcal{E}   (   \mathcal{M},
  \mathcal{H}   )$   such   that   the   prior   probability
  distribution   $\Tr[  \boldsymbol{\rho}   (  \cdot   )]  =
  |\mathcal{M}|^{-1}$  is  uniform, there  exists  numbering
  $\mathbf{n}^*  \in  \mathcal{N}_{\mathcal{M}}$  such  that
  measurement          $\boldsymbol{\pi}_{\boldsymbol{\rho},
    \mathbf{n}^*}$ minimizes the guesswork, that is
  \begin{align*}
    G_{\textrm{min}}  \left( \boldsymbol{\rho}  \right) =  G
    \left(                                \boldsymbol{\rho},
    \boldsymbol{\pi}_{\boldsymbol{\rho},       \mathbf{n}^*}
    \right).
  \end{align*}
\end{cor}

We remark that Corollary~\ref{cor:qubit} recasts the minimum
guesswork  problem, by  definition  an optimization  problem
over a  \textit{continuous} set, as an  optimization problem
over a \textit{finite} set,  that can be therefore performed
by exhaustive search.

\begin{proof}
  Since by hypothesis $\Tr [ \boldsymbol{\rho} ( \cdot ) ] =
  | \mathcal{M} |^{-1}$, one has
  \begin{align*}
    \Tr   \left[  E_{\boldsymbol{\rho}}   \left(  \mathbf{n}
      \right) \right] = 0,
  \end{align*}
  for   any   $\mathbf{n}  \in   \mathcal{N}_{\mathcal{M}}$.
  Hence,    since    by    hypothesis    $\mathcal{H}$    is
  two-dimensional, one has
  \begin{align*}
    \left|  E_{\boldsymbol{\rho}} \left(  \mathbf{n} \right)
    \right|    =   \left\|    E_{\boldsymbol{\rho}}   \left(
    \mathbf{n} \right) \right\|_1 \frac{\openone}2,
  \end{align*}
  for any $\mathbf{n} \in \mathcal{N}_{\mathcal{M}}$. Hence,
  the range $\left| E_{\boldsymbol{\rho}} \left( \mathcal{N}
  \left(  \mathcal{M} \right)  \right)  \right|$ is  totally
  ordered.  Hence, there exists $\mathbf{n}^*$ such that
  \begin{align*}
    \left| E_{\boldsymbol{\rho}} \left( \mathbf{n}^* \right)
    \right|   \ge    \left|   E_{\boldsymbol{\rho}}   \left(
    \mathbf{n}  \right)  \right|  \ge  E_{\boldsymbol{\rho}}
    \left( \mathbf{n} \right),
  \end{align*}
  for any $\mathbf{n} \in \mathcal{N}_{\mathcal{M}}$.  Hence
  the statement follows from Theorem~\ref{thm:helstrom}.
\end{proof}

\section{Explicit Examples}
\label{sec:examples}

In  this section  we provide  the minimum  guesswork of  any
qubit regular polygonal or polyhedral ensemble by explicitly
solving  the  optimization  over   a  finite  set  given  by
Corollary~\ref{cor:qubit}.

\begin{cor}[Regular polygonal ensembles]
  \label{cor:polygons}
  For  any discrete  set $\mathcal{M}$,  any two-dimensional
  Hilbert  space $\mathcal{H}$,  and any  bijective ensemble
  $\boldsymbol{\rho}     \in    \mathbb{M}(     \mathcal{M},
  \mathcal{H} )$ whose range $\boldsymbol{\rho}( \mathcal{M}
  )$  is proportional  to  a regular  polygon  in the  Bloch
  circle, one has
  \begin{align*}
    G_{\textrm{min}}  \left(   \boldsymbol{\rho}  \right)  =
    \frac{\left|    \mathcal{M}    \right|     +    1}2    -
    \frac12  \begin{cases}   \frac{2  \sqrt{3   \cos  \left(
          \frac{\pi}{\left| \mathcal{M} \right|} \right)^2 +
          1}}{\left|   \mathcal{M}   \right|   \sin   \left(
        \frac{\pi}{\left| \mathcal{M} \right|} \right)^2}, &
      \textrm{if $|\mathcal{M}|$ even,}\\ \frac{ \cos \left(
        \frac{\pi}{2 \left| \mathcal{M} \right|} \right) } {
        \left| \mathcal{M} \right|  \sin \left( \frac{\pi}{2
          \left|   \mathcal{M}    \right|}   \right)^2},   &
      \textrm{if $|\mathcal{M}|$ odd.}
      \end{cases}
  \end{align*}
\end{cor}

\begin{figure}[htb!]
  \begin{center}
    \includegraphics[width=\columnwidth]{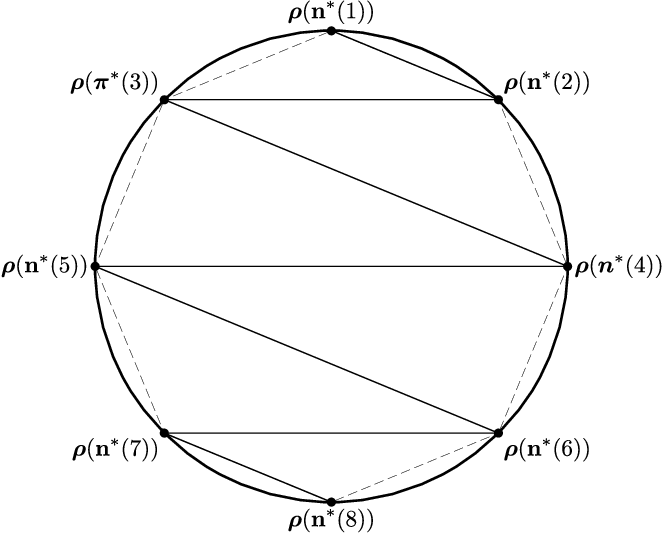}
  \end{center}
  
  \caption{The    figure     illustrates    the    numbering
    $\mathbf{n}^* \in \mathcal{N} ( \mathcal{M} )$ such that
    $G_{\textrm{min}}   (   \boldsymbol{\rho}   )  =   G   (
    \boldsymbol{\rho},  \boldsymbol{\pi}_{\boldsymbol{\rho},
      \mathbf{n}^*})$,    when     $\boldsymbol{\rho}    \in
    \mathcal{E}   (  \mathcal{M},   \mathbb{R}^2  )$   is  a
    bijective          ensemble           such          that
    $\boldsymbol{\rho}(\mathcal{M})$  is  proportional to  a
    regular polygon  ($|\mathcal{M}| = 8$ in  the figure) in
    the  Bloch circle.}

  \label{fig:polygons}
\end{figure}

\begin{proof}
  Due to  Corollary~\ref{cor:qubit}, there  exists numbering
  $\mathbf{n}^* \in  \mathcal{N} ( \mathcal{M} )$  such that
  $G_{\textrm{min}}       (\boldsymbol{\rho})      =       G
  (\boldsymbol{\rho},   \boldsymbol{\pi}_{\boldsymbol{\rho},
    \mathbf{n}^*})$.   Due   to  Lemma~\ref{lmm:decreasing},
  $q_{\boldsymbol{\rho},
    \boldsymbol{\pi}_{\boldsymbol{\rho},  \mathbf{n}^*}}$ is
  not increasing.  One way of representing $\mathbf{n}^*$ is
  as follows.  Without loss  of generality take $\mathcal{M}
  =    \{   1,    \dots,   |    \mathcal{M}   |    \}$   and
  $\boldsymbol{\rho}(m)  =  |\mathcal{M}|^{-1}  \ket{\psi_m}
  \!\!  \bra{\psi_m}$, where $\ket{\psi_m} =  \cos ( 2 \pi m
  /  |  \mathcal{M} |  )  \ket{0}  + \sin  (  2  \pi m  /  |
  \mathcal{M} | ) \ket{1}$.  Then one has
  \begin{align*}
    \mathbf{n}^*  \left( m  \right)  = \begin{cases}  2 m  &
      \textrm{ if } m < \frac{\left| \mathcal{M} \right|}2 +
      \frac14,\\  -2m +2  \left|  \mathcal{M}  \right| +  1&
      \textrm{ otherwise.}
    \end{cases}
  \end{align*}
  Numbering     $\mathbf{n}^*$     is     illustrated     in
  Fig.~\ref{fig:polygons}  for $|  \mathcal{M} |  = 8$.   By
  summing finite  trigonometric series,  for $|\mathcal{M}|$
  even one has
  \begin{align*}
    E_{\boldsymbol{\rho}}  \left(   \mathbf{n}^*  \right)  =
    \frac1{\left|  \mathcal{M}  \right|} \begin{bmatrix}  -2
      \cot  \left(  \frac{\pi}{\left|  \mathcal{M}  \right|}
      \right)^2  -1   &  -  \cot   \left(  \frac{\pi}{\left|
        \mathcal{M}   \right|}  \right)\\   -  \cot   \left(
      \frac{\pi}{\left|  \mathcal{M}  \right|} \right)  &  2
      \cot  \left(  \frac{\pi}{\left|  \mathcal{M}  \right|}
      \right)^2 +1
    \end{bmatrix},
  \end{align*}
  and for $|\mathcal{M}|$ odd one has
  \begin{align*}
    E_{\boldsymbol{\rho}}  \left(   \mathbf{n}^*  \right)  =
    \frac1{2 \left|  \mathcal{M} \right|}  \begin{bmatrix} -
      \cot \left(  \frac{\pi}{2 \left|  \mathcal{M} \right|}
      \right)^2   &  -   \cot  \left(   \frac{\pi}{2  \left|
        \mathcal{M}   \right|}  \right)\\   -  \cot   \left(
      \frac{\pi}{2  \left|  \mathcal{M} \right|}  \right)  &
      \cot \left(  \frac{\pi}{2 \left|  \mathcal{M} \right|}
      \right)^2
    \end{bmatrix}.
  \end{align*}
  By explicit computation one has
  \begin{align*}
    \left\|   E_{\boldsymbol{\rho}}    \left(   \mathbf{n}^*
    \right) \right\|_1 =  \begin{cases} 2 \frac{\sqrt{3 \cos
          \left(   \frac{\pi}{\left|  \mathcal{M}   \right|}
          \right)^2  + 1}}{\left|  \mathcal{M} \right|  \sin
        \left(   \frac{\pi}{\left|    \mathcal{M}   \right|}
        \right)^2},     &     \textrm{if     $|\mathcal{M}|$
        even,}\\  \frac{  \cos  \left(  \frac{\pi}{2  \left|
          \mathcal{M}   \right|}   \right)    }   {   \left|
        \mathcal{M} \right| \sin  \left( \frac{\pi}{2 \left|
          \mathcal{M}  \right|}   \right)^2},  &  \textrm{if
        $|\mathcal{M}|$ odd.}
      \end{cases}
  \end{align*}
  Hence the statement follows from Eq.~\eqref{eq:mingw}.
\end{proof}

\begin{cor}[Regular polyhedral ensembles]
  \label{cor:polyhedra}
  For  any discrete  set $\mathcal{M}$,  any two-dimensional
  Hilbert  space $\mathcal{H}$,  and any  bijective ensemble
  $\boldsymbol{\rho}   \in    \mathcal{E}   (   \mathcal{M},
  \mathcal{H}   )$   whose    range   $\boldsymbol{\rho}   (
  \mathcal{M} )$ is proportional  to a regular polyhedron in
  the Bloch sphere, one has
  \begin{align*}
    G_{\textrm{min}}   \left(    \boldsymbol{\rho}   \right)
    = \begin{cases} \frac52 - \frac{\sqrt{15}}{6} \sim 1.9 &
      \textrm{if  $|  \mathcal{M}  |   =  4$,}\\  \frac72  -
      \frac{\sqrt{35}}6 \sim 2.5 & \textrm{if $| \mathcal{M}
        | =  6$,}\\ \frac92  - \frac{\sqrt{7}}2  \sim 3.2  &
      \textrm{if  $| \mathcal{M}  |  =  8$,}\\ \frac{13}2  -
      \frac{\sqrt{110 \left( 65  + 29 \sqrt{5} \right)}}{60}
      \sim   4.5   &   \textrm{if   $|   \mathcal{M}   |   =
        12$,}\\  \frac{21}2 -  \frac{\sqrt{6  \left( 3321  +
          1483 \sqrt{5} \right) }}{60} \sim 7.2 & \textrm{if
        $| \mathcal{M} | = 20$.}
    \end{cases}
  \end{align*}
\end{cor}

\begin{figure}[htb!]
  \begin{center}
    \includegraphics[width=\columnwidth]{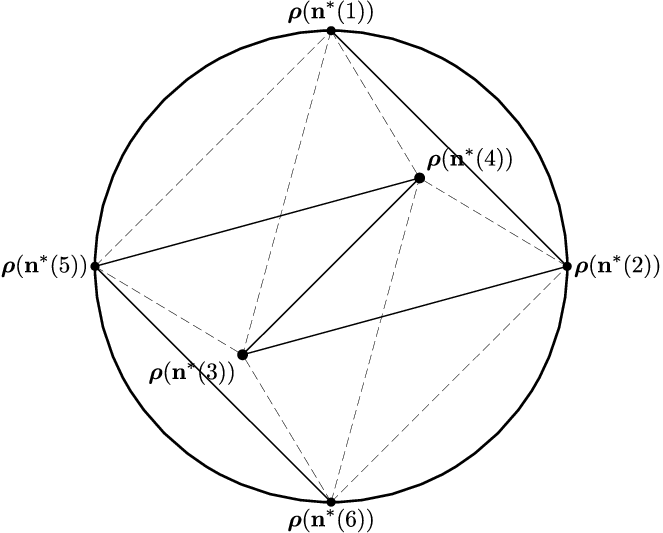}
  \end{center}
  
  \caption{The    figure     illustrates    the    numbering
    $\mathbf{n}^* \in \mathcal{N} ( \mathcal{M} )$ such that
    $G_{\textrm{min}}   (   \boldsymbol{\rho}   )  =   G   (
    \boldsymbol{\rho}, \boldsymbol{\pi}_{ \boldsymbol{\rho},
      \mathbf{n}^*})$,    when     $\boldsymbol{\rho}    \in
    \mathcal{E}   (  \mathcal{M},   \mathbb{C}^2  )$   is  a
    bijective          ensemble           such          that
    $\boldsymbol{\rho}(\mathcal{M})$  is  proportional to  a
    regular polyhedron  ($|\mathcal{M}| = 6$ in  the figure)
    in the  Bloch sphere.}

  \label{fig:polyhedra}
\end{figure}

\begin{proof}
  Due to  Corollary~\ref{cor:qubit}, there  exists numbering
  $\mathbf{n}^* \in  \mathcal{N} ( \mathcal{M} )$  such that
  $G_{\textrm{min}}   (   \boldsymbol{\rho}    )   =   G   (
  \boldsymbol{\rho},    \boldsymbol{\pi}_{\boldsymbol{\rho},
    \mathbf{n}^*})$.    For  $|   \mathcal{M}|   =  4$   any
  $\mathbf{n}^* \in \mathcal{N}(  \mathcal{M})$ is such that
  $G_{\textrm{min}}   (   \boldsymbol{\rho}    )   =   G   (
  \boldsymbol{\rho},    \boldsymbol{\pi}_{\boldsymbol{\rho},
    \mathbf{n}^*})$, hence the result for $| \mathcal{M} | =
  4$ follows.   Let us consider  the case $|  \mathcal{M}| >
  4$.        Due        to       Lemma~\ref{lmm:decreasing},
  $q_{\boldsymbol{\rho},
    \boldsymbol{\pi}_{\boldsymbol{\rho},  \mathbf{n}^*}}$ is
  not  increasing.  Since  the  range  $\boldsymbol{\rho}  (
  \mathcal{M} )$ is centrally symmetric, that is
  \begin{align*}
    \boldsymbol{\rho}  \left( \mathcal{M}  \right) =  \left|
    \mathcal{M}  \right|^{-1}  \openone -  \boldsymbol{\rho}
    \left( \mathcal{M} \right),
  \end{align*}
  any     $\mathbf{n}^*$     with     $q_{\boldsymbol{\rho},
    \boldsymbol{\pi}_{\boldsymbol{\rho}, \mathbf{n}^*}}$ not
  increasing satisfies
  \begin{align*}
    \boldsymbol{\rho}   \left(  \mathbf{n}^*   \left(  \cdot
    \right)     \right)    +     \boldsymbol{\rho}    \left(
    \overline{\mathbf{n}}^* \left(  \cdot \right)  \right) =
    \left| \mathcal{M} \right|^{-1} \openone.
  \end{align*}
  Since fixing the value of $\mathbf{n}^*(t)$ also fixes the
  value    of     $\overline{\mathbf{n}}^*(t)$,    numbering
  $\mathbf{n}^*$ can be found  in $| \mathcal{M} |!!$ steps.
  Also, since  regular polyhedra are vertex  transitive, the
  choice  of  $\mathbf{n}^*  (  1 )$  is  irrelevant,  hence
  $\mathbf{n}^*$ can  be found  in $|  \mathcal{M} -  2 |!!$
  steps.  The  exhaustive search  is practical even  for the
  dodecahedron for which $| \mathcal{M} | = 20$ and hence $|
  \mathcal{M} - 2 |!!  \sim 10^8$.  Numbering $\mathbf{n}^*$
  is   illustrated   in  Fig.~\ref{fig:polyhedra}   for   $|
  \mathcal{M} | = 6$. Hence the results for $| \mathcal{M} |
  > 4$   follow.    Further   details  can   be   found   in
  Ref.~\cite{DBK21},  where  algorithms  for  the  classical
  computation of the quantum  guesswork in analytical closed
  form  based  on  the  present  results  are  provided  and
  analyzed.
\end{proof}

\section*{Appendix}

In this appendix we derive  technical results needed for the
derivation of our main results.

\begin{lmm}
  \label{lmm:gw}
  For any  finite set $\mathcal{M}$, any  finite dimensional
  Hilbert     space      $\mathcal{H}$,     any     ensemble
  $\boldsymbol{\rho}    \in     \mathcal{E}    (\mathcal{M},
  \mathcal{H})$,   and   any  numbering-valued   measurement
  $\boldsymbol{\pi}         \in        \mathcal{P}         (
  \mathcal{N}_{\mathcal{M}}, \mathcal{H})$, the guesswork $G
  ( \boldsymbol{\rho}, \boldsymbol{\pi})$ is given by
  \begin{align*}
    G \left(  \boldsymbol{\rho}, \boldsymbol{\pi}  \right) =
    \frac{\left|  \mathcal{M}   \right|  +  1}2   +  \frac12
    \sum_{\mathbf{n}   \in  \mathcal{N}_{\mathcal{M}}}   \Tr
    \left[  E_{\boldsymbol{\rho}} \left(  \mathbf{n} \right)
      \frac{\boldsymbol{\pi}  \left(  \mathbf{n}  \right)  -
        \boldsymbol{\pi}     \left(    \overline{\mathbf{n}}
        \right)}2 \right].
  \end{align*}
\end{lmm}

\begin{proof}
  By definition of map  $E_{\boldsymbol{\rho}}$ one has $G (
  \boldsymbol{\rho}, \boldsymbol{\pi} ) = (| \mathcal{M} | +
  1 + x_{\boldsymbol{\rho}, \boldsymbol{\pi}}) / 2$, where
  \begin{align*}
    x_{\boldsymbol{\rho},        \boldsymbol{\pi}}        :=
    \sum_{\mathbf{n}   \in  \mathcal{N}_{\mathcal{M}}}   \Tr
    \left[  E_{\boldsymbol{\rho}} \left(  \mathbf{n} \right)
      \boldsymbol{\pi} \left( \mathbf{n} \right) \right].
  \end{align*}
  Using the identity $E_{\boldsymbol{\rho}} ( \mathbf{n} ) =
  -E_{\boldsymbol{\rho}} ( \overline{\mathbf{n}} )$ one has
  \begin{align*}
    x_{\boldsymbol{\rho},         \boldsymbol{\pi}}        =
    \sum_{\mathbf{n}   \in  \mathcal{N}_{\mathcal{M}}}   \Tr
    \left[  E_{\boldsymbol{\rho}} \left(  \mathbf{n} \right)
      \frac{\boldsymbol{\pi}  \left(  \mathbf{n}  \right)  -
        \boldsymbol{\pi}     \left(    \overline{\mathbf{n}}
        \right) }2 \right].
  \end{align*}
  Hence the statement follows.
\end{proof}
  
For any  finite dimensional Hilbert space  $\mathcal{H}$ and
any  operator  $A  \in  \mathcal{L} (  \mathcal{H})  $,  let
$\mathcal{P}_A :  \mathcal{L} (\mathcal{H})  \to \mathcal{L}
(\mathcal{H})$ be a dephasing  map given by $\mathcal{P}_A (
\cdot  )  =  \sum_a   \bra{a}  \cdot  \ket{a}  \ket{a}  \!\!
\bra{a}$,  where  $\{  \ket{a}  \}$ is  a  complete  set  of
eigenvectors of $A$.

\begin{lmm}
  \label{lmm:majorization1}
  For any finite dimensional Hilbert space $\mathcal{H}$ and
  any $X,  A \in \mathcal{L}  (\mathcal{H})$, if $| X  | \le
  \openone$ one has that $| \Tr [ A X ] | \le \| A \|_1$.
\end{lmm}

\begin{proof}
  Since $\mathcal{P}_A$ is linear,  positive, and unital, by
  the hypothesis it follows that $| \mathcal{P}_A ( X) | \le
  \openone$.  Since $\Tr [ A X ] = \Tr [ A \mathcal{P}_A ( X
    ) ]$, the statement follows.
\end{proof}

\begin{lmm}
  \label{lmm:majorization2}
  For any finite dimensional Hilbert space $\mathcal{H}$ and
  any $X, A \in \mathcal{L} (\mathcal{H})$, if $-X \le A \le
  X$ one has that $\| A \|_1 \le \| X \|_1$.
\end{lmm}

\begin{proof}
  Since   $\mathcal{P}_A$  is   linear   and  positive   and
  $\mathcal{P}_A (A) = A$, by the hypothesis it follows that
  $-  \mathcal{P}_A (  X) \le  A \le  \mathcal{P}_A (  X )$.
  Since $[ \mathcal{P}_A (X), A]  = 0$ and by the hypothesis
  it  follows  that   $X  \ge  0$,  one  has  $|   A  |  \le
  \mathcal{P}_A  ( X  )$.   Since  $\mathcal{P}_A$ is  trace
  preserving, by tracing both sides the statement follows.
\end{proof}

The following  lemma provides a necessary  condition for any
measurement to attain the minimum guesswork.

\begin{lmm}
  \label{lmm:decreasing}
  For any discrete set $\mathcal{M}$, any finite dimensional
  Hilbert    space   $\mathcal{H}$,    and   any    ensemble
  $\boldsymbol{\rho}    \in     \mathcal{E}    (\mathcal{M},
  \mathcal{H})$,   a   measurement   $\boldsymbol{\pi}   \in
  \mathcal{P}    (\mathcal{N}_{\mathcal{M}},   \mathcal{H})$
  minimizes  the  guesswork,  that  is  $G_{\textrm{min}}  (
  \boldsymbol{\rho}    )    =   G    (    \boldsymbol{\rho},
  \boldsymbol{\pi}   )$,   only  if   $p_{\boldsymbol{\rho},
    \boldsymbol{\pi}} (\mathbf{n}, \cdot)$ is not increasing
  for any $\mathbf{n} \in \mathcal{N}( \mathcal{M} )$.
\end{lmm}

\begin{proof}
  We  show that  for any  measurement $\boldsymbol{\pi}  \in
  \mathcal{P}(\mathcal{N}_{\mathcal{M}}, \mathcal{H})$ there
  exists     a     measurement    $\boldsymbol{\pi}'     \in
  \mathcal{P}(\mathcal{N}_{\mathcal{M}},  \mathcal{H})$ such
  that       $p_{\boldsymbol{\rho},       \boldsymbol{\pi}'}
  (\mathbf{n}, \cdot)$ is not increasing for any $\mathbf{n}
  \in \mathcal{N}_{\mathcal{M}}$  and $G( \boldsymbol{\rho},
  \boldsymbol{\pi}'    )     \le    G(    \boldsymbol{\rho},
  \boldsymbol{\pi}  )$,   with  equality  if  and   only  if
  $p_{\boldsymbol{\rho},    \boldsymbol{\pi}}   (\mathbf{n},
  \cdot)    =    p_{\boldsymbol{\rho},    \boldsymbol{\pi}'}
  (\mathbf{n},    \cdot)$    for   any    $\mathbf{n}    \in
  \mathcal{N}_{\mathcal{M}}$.  Let $\{ g_{\mathbf{n}} : \{1,
  \dots, | \mathcal{M} | \}  \to \{1, \dots, | \mathcal{M} |
  \}   \;   |    \;   g_{\mathbf{n}}   \textrm{   bijective}
  \}_{\mathbf{n} \in \mathcal{N}_{\mathcal{M}}}$  be a family
  of    permutations   such    that   $p_{\boldsymbol{\rho},
    \boldsymbol{\pi}}  (  \mathbf{n}, g_{\mathbf{n}}(  \cdot
  ))$   is   not   increasing  for   any   $\mathbf{n}   \in
  \mathcal{N}_{      \mathcal{M}}$.       Let      $f      :
  \mathcal{N}_{\mathcal{M}}  \to  \mathcal{N}_{\mathcal{M}}$
  be given by
  \begin{align*}
    f   \left(  \mathbf{n}   \right)  :=   \mathbf{n}  \circ
    g_{\mathbf{n}},
  \end{align*}
  for any  $\mathbf{n} \in  \mathcal{N}_{\mathcal{M}}$.  Let
  $\boldsymbol{\pi}'             \in            \mathcal{P}(
  \mathcal{N}_{\mathcal{M}}, \mathcal{H}  ) $ be  the coarse
  graining of $\boldsymbol{\pi}$ given by
  \begin{align*}
    \boldsymbol{\pi}'   \left(    \mathbf{n}'   \right)   :=
    \sum_{\mathbf{n}  \in f^{-1}\left[  \mathbf{n}' \right]}
    \boldsymbol{\pi} \left( \mathbf{n} \right),
  \end{align*}
  for any $\mathbf{n}' \in \mathcal{N}_{\mathcal{M}}$, where
  $f^{-1}  [ \mathbf{n}'  ]$  denotes  the counter-image  of
  $\mathbf{n}'$   with   respect   to  $f$.    By   explicit
  computation one has
  \begin{align*}
    q_{\boldsymbol{\rho},   \boldsymbol{\pi}'}    \left(   t
    \right)       &       =      \sum_{\mathbf{n}'       \in
      \mathcal{N}_{\mathcal{M}}} \sum_{\mathbf{n} \in f^{-1}
      \left[     \mathbf{n}'     \right]}     \Tr     \left[
      \boldsymbol{\rho} \left( \mathbf{n}'  \left( t \right)
      \right)  \boldsymbol{\pi}  \left(  \mathbf{n}  \right)
      \right]     \\     &    =     \sum_{\mathbf{n}     \in
      \mathcal{N}_{\mathcal{M}}}          \Tr         \left[
      \boldsymbol{\rho}  \left(  f \left(\mathbf{n}  \right)
      \left(  t  \right)   \right)  \boldsymbol{\pi}  \left(
      \mathbf{n} \right) \right] \\ & = \sum_{\mathbf{n} \in
      \mathcal{N}_{\mathcal{M}}}       p_{\boldsymbol{\rho},
      \boldsymbol{\pi}}  \left(  \mathbf{n},  g_{\mathbf{n}}
    \left( t \right) \right),
  \end{align*}
  for any $t \in \{1, \dots,  | \mathcal{M} | \}$.  Hence by
  construction
  \begin{align*}
    \sum_{t    \in   \left\{    1,   \dots,    T   \right\}}
    q_{\boldsymbol{\rho},   \boldsymbol{\pi}'}    \left(   t
    \right) \ge  \sum_{t \in  \left\{ 1, \dots,  T \right\}}
    q_{\boldsymbol{\rho}, \boldsymbol{\pi}} (t)
  \end{align*}
  for any  $T \in  \{ 1,  \dots, |  \mathcal{M} |  \}$, with
  equality   if    and   only    if   $p_{\boldsymbol{\rho},
    \boldsymbol{\pi}}       (\mathbf{n},      \cdot)       =
  p_{\boldsymbol{\rho},    \boldsymbol{\pi}'}   (\mathbf{n},
  \cdot)$ for any $\mathbf{n}  \in \mathcal{N} ( \mathcal{M}
  )$.   Hence   the  statement  follows  by   definition  of
  guesswork.
\end{proof}

\section{Conclusion}
\label{sec:conclusion}

The guesswork  of a quantum ensemble  quantifies the minimum
number of guesses  needed in average to  correctly guess the
state of the ensemble, when only one state can be queried at
a time.  Here, we derived  analytical solutions subject to a
finite set of conditions, including analytical solutions for
any  qubit ensemble  with uniform  probability distribution,
thus disproving the conjecture~\cite{CCWF15} that analytical
solutions only exist for binary and symmetric ensembles.  As
explicit examples,  we computed the guesswork  for any qubit
regular polygonal  and polyhedral  ensemble, thus  proving a
conjecture~\cite{HKDW20}  on  the  guesswork of  the  square
qubit ensemble.

\section{Acknowledgments}

M.~D. is  grateful to  Eric Hanson  and Nilanjana  Datta for
insightful discussions  during a visit to  the University of
Cambridge.   M.~D.  acknowledges  support from  MEXT Quantum
Leap    Flagship   Program    (MEXT   Q-LEAP)    Grant   No.
JPMXS0118067285, JSPS  KAKENHI Grant Number  JP20K03774, and
the  International  Research  Unit of  Quantum  Information,
Kyoto  University. F.   B.  acknowledges  support from  MEXT
Quantum  Leap  Flagship  Program  (MEXT  Q-LEAP)  Grant  No.
JPMXS0120319794;    from    MEXT-JSPS    Grant-in-Aid    for
Transformative Research Areas  (A) ``Extreme Universe'', No.
21H05183;  from  JSPS  KAKENHI   Grants  No.   19H04066  and
20K03746. T.~K. acknowledges support  from MEXT Quantum Leap
Flagship  Program (MEXT  Q-LEAP) Grant  No.  JPMXS0118067285
and  No.   JPMXS0120319794  and JSPS  KAKENHI  Grant  Number
16H01705, 17H01695, 19K22849 and 21H04879.

\textbf{Michele Dall'Arno}  received his PhD  in theoretical
physics from  the University  of Pavia,  Italy, in  2012. He
held   post-doc  positions   in  ICFO   (Barcelona),  Nagoya
University   (Japan),  and   the   National  University   of
Singapore.  Since 2020, he is assistant professor in quantum
information in Kyoto University,  and visiting researcher in
Waseda University, Tokyo.

\textbf{Francesco Buscemi} received his PhD in theoretical physics from the University of Pavia, Italy, in 2006. After post-doc positions in Tokyo, Japan, and Cambridge, UK, he joined Nagoya University in 2009, where he is a professor in the department of mathematical informatics. In 2018 he was awarded the Birkhoff-von Neumann Prize by the International Quantum Structures Association.

\textbf{Takeshi Koshiba}  received his  PhD degree  from the
Tokyo Institute  of Technology.  He  is a full  professor at
the  Department of  Mathematics,  Faculty  of Education  and
Integrated Arts and Sciences,  Waseda University, Japan. His
interests  include  theoretical  and  applied  cryptography,
randomness   in    algorithms,   and    quantum   computing,
cryptography and information.

\end{document}